\documentclass[aps,prd,groupedaddress,showpacs,showkeys,nofootinbib]{revtex4}
\usepackage{graphicx,amssymb,amsmath}
\usepackage[english]{babel}

\begin{document}

\title{Physical non-equivalence of the Jordan and Einstein frames}

\author{S.Capozziello${}^{1,2}$, P. Martin-Moruno${}^{3}$, C. Rubano${}%
^{1,2}$} \affiliation{} \affiliation{${}^{1}$ Dipartimento
di Scienze Fisiche, Universit\`{a} di Napoli "Federico II" and
${}^{2}$INFN Sez. di Napoli, Compl. Univ. Monte S. Angelo, Ed.N, Via
Cinthia, I-80126 Napoli, Italy,} \affiliation{${}^{3}$ Colina
de los Chopos, Instituto de Fisica Fundamental, Consejo Superior
de Investigaciones Cientificas, Serrano 121, 28006 Madrid, Spain.}
\date{\today}

\begin{abstract}
We show, considering a specific $f(R)$-gravity model, that the Jordan frame and the
Einstein frame could be physically non-equivalent, although they are
connected by a conformal transformation which yields a
mathematical equivalence. Calculations are performed
analytically and this non-equivalence is shown in an unambiguous way. 
However this statement strictly depends on the considered physical quantities that have to be carefully selected.
\end{abstract}


\maketitle

\section{Introduction}\label{intro}
The current accelerated expansion of the Universe, supported by a
large number of observational data \cite{uno,dos,tres}, is one of
the most challenging issues of  the modern physics. The assumption that
 Einstein's General Relativity (GR) is the correct theory of
gravity leads to the consideration that approximately the $70\%$
of the energy density of the universe should be an unknown form of fluid
called ``dark energy'', responsible of the mentioned acceleration.
Even more, the largest part of the matter content 
is not consituted  by  standard baryonic matter, but by another unknown component
called ``cold dark matter'' (CDM), constituting  about the $25\%$ of
the total matter-energy budget. The most popular model able to
describe this scenario is the $\Lambda$CDM, where it is considered
that the dark energy component is simply the cosmological
constant. Although $\Lambda$CDM model fits to a wide range of data
\cite{Seljak:2004xh}, it is affected by strong theoretical
shortcomings \cite{Carroll:1991mt}. Specifically there is the
cosmological constant problem \cite{Gold}, regarding the fact that
the predicted value of the quantum-field vacuum energy density and
the observed cosmological value are currently separated by 120 orders of
magnitude, or the cosmic coincidence problem, which opens the
question about why the  today observed  values of  the CDM density 
and the cosmological constant energy density are of the same order of magnitude.

These shortcomings have motivated the study of a plethora of
models which consider a dynamical dark energy characterized by an
equation of state parameter $w<-1/3$ ($w=p/\rho$), recovering the
$\Lambda$CDM model in the particular case that one has a constant
parameter $w=-1$. Although these models could avoid in some cases
the mentioned problems, \cite{Chimento:2003iea}, the origin of
this  fluid which produces anti-gravitational effects,
violating at least one of the energy conditions \cite{Visser}, remains a
mystery.

On the other hand, up to now, there is no definitive candidate
for CDM, in spite of the efforts to identify its particle nature
by its non-gravitational effects from space and ground-based experiments (comments in some experimental
programs can be find in \cite{Hooper:2008sn} and one of the main goal of the Large Hadron Collider at CERN is the identification of these particles \cite{LHC}).

Since the validity on large astrophysical and cosmological scales
of GR has never been tested, one could  suppose that current
observational datasets  imply the non-validity of GR at those scales.
Therefore, {\it Extended Theories of Gravity} (ETGs), which was
initially introduced by  quantum motivations, have been taken
seriously in consideration. ETGs modify and enlarge the Einstein
theory, adding into the effective action physically motivated
higher order curvature invariants and/or non-minimally coupled
scalar fields \cite{Buchbinder:1992rb,Ruggiero}.

Among ETGs,  $f(R)$-theories are becoming of great
interest, since they are the minimal extension of GR able to match
the data without need of any dark energy or dark matter \cite{jcap}. These
theories modify the Einstein's action including a generic function
$f(R)$ of the Ricci scalar $R$ instead of rigidly considering the Hilbert-Einstein action linear in $R$ \cite{odin,francaviglia,defelice,faraoni}.

The Hilbert-Einstein action and the 
$f(R)$-action can be related by a conformal transformation
\cite{Allemandi, Capozziello:2005mj, OdiTroisi}, being the corresponding equations also
connected by the same transformation. This fact shows that the
Einstein frame and the Jordan frame are mathematically equivalent \cite{magnano} 
but they could not be physically equivalent as pointed out in several papers (see e.g \cite{OdiTroisi,Mercadante,Clifton}).

This is an old argument widely discussed in last decades
 (see e.g. \cite{vassi} where a detailed discussion for dilaton gravity in two dimensions is reported).
In \cite{mariusz1}, for example, the  problem of physical non-equivalence of  
conformal frames has been considered: in that case 
 the work done by a conformal transformation is capable of  "creating" 
matter and so the two frames have not the same physical 
meaning (one is empty and another has matter). Anyway, this could mean
that the conformal transformations change physics unlike the coordinate 
transformations.
Besides,  the method of the conformal transformation  can be used to study 
the problem of energy-momentum content of the gravitational field using 
statefinders \cite{mariusz2}.

This is an open question that, up to now, has not been completely solved (see \cite{faraoni2} for a review on the topic).
In particular,  a strong debate has been pursued about the Newtonian limit (i.e. small velocity and weak
field) of fourth order gravity models. According to some authors, the Newtonian limit of $f (R)$-gravity
is equivalent to the one of BransÐDicke gravity with the Brans-Dicke parameter $\omega = 0$, so that the PPN parameters of these
models turn out to be ill-defined. In a recent paper \cite{Arturo}, this point has been carefully  discussed  considering that fourth
order gravity models are dynamically equivalent to the OÕHanlon Lagrangian \cite{ohanlon}. This is a special case of
scalarÐtensor gravity characterized only by self-interaction potential and that, in the Newtonian limit,
this implies a non-standard behavior that cannot be compared with the usual PPN limit of GR. The result turns out to be completely different from the one of BransÐDicke theory and, in
particular,  suggests that it is misleading to consider the PPN parameters of this theory with $\omega = 0$ in
order to characterize the homologous quantities of $f (R)$-gravity.  In other words, this result can be considered an indication of the fact that conformally transformed theories could  $not$ be physically equivalent (see e.g. \cite{Arturo}). However, this statement has to be supported by the fact that methods to measure  observable physical quantities should be completely independent of the frames or, at least, the relation of their observed values into the frames well established.  
 
The aim of this work is to prove that the physical non-equivalence of
Jordan frame and Einstein frame could be exactly demonstrated considering a suitable model and selecting physically reliable quantities.
 For this reason, we will take into account a
$f(R)$-model which allows us to compare  analytically the two frames showing the physical differences. 

 The layout of this Letter is the following. 
 In Sec.~\ref{uno},  we review the $f(R)$- cosmological model
presented in \cite{Capozziello:2008im}. It is particularly interesting being exactly integrable 
and capable of
describing dust matter (decelerated) phase and the following dark energy (accelerated) phase under the same standard.
 In Sec.~\ref{dos},  we perform the  conformal transformation 
obtaining  the mathematically  equivalent model in the
Einstein frame. The comparison of the  model,  in Jordan's and Einstein's frame,  is presented in
Sec.~\ref{tres} showing the possible physical non-equivalnce. Being the calculations  completely
analytical,  the comparison can be perform in an unambiguous way.
Finally, in Sec.~\ref{cuatro}, we summarize the results and draw our  conclusions.

\section{The model}\label{uno}
A general action describing $f(R)$- gravity in four dimensions  is 
\begin{equation}\label{Ag}
{\cal A}=\int d^4 x\,\sqrt{-g}\,f(R)\,+{\cal A}_m\, ,
\end{equation}
where $f(R)$ is a generic function of the Ricci scalar $R$ and
${\cal A}_m$ is the action of a perfect fluid minimally coupled
with gravity. Obviously assuming $f(R)=R$ the standard Einstein
theory is recovered. Varying with respect to $g_{\mu\nu}$, we get the field equations
\begin{equation}\label{eqeinsteinJ}
G_{\mu\nu}=T_{\mu\nu}^{curv}+\frac{T_{\mu\nu}^{m}}{2f'(R)}\, ,
\end{equation}
where
\begin{equation}
G_{\mu\nu}=R_{\mu\nu}-\frac{1}{2}R\,g_{\mu\nu}
\end{equation}
and $T_{\mu\nu}^{curv}$ is an effective stress-energy tensor
constructed by curvature terms in the following way
\begin{equation}
T_{\mu\nu}^{curv}=\frac{1}{f'(R)}\left\{\frac{1}{2}g_{\mu\nu}\left[f(R)-R\,f'(R)\right]
+ f'(R)_{;\mu\nu}-g_{\mu\nu}f'(R)_{;\alpha}^{;\alpha}\right\}\, .
\end{equation}
This tensor is zero for $f(R)=R$. The prime indicates derivatives with respect to $R$.

In a Friedmann-Robertson-Walker (FRW) metric, taking into account
a dust-matter perfect fluid, a point-like Lagrangian can be
obtained
\begin{equation}\label{pL}
{\cal L}= a^3\,\left[f(R)-f'(R)\,R\right]+6\,a^2\,f''(R)\,\dot
R\,\dot a +6\,f'(R)\,a\,\dot a^2-6k\,f'(R)\,a+D\, ,
\end{equation}
where $D$ represents the standard amount of dust fluid, such that
$\rho=D/a^3$ \cite{C2}. The  energy function $E_{{\cal L}}$,
 corresponding to the $\{0,0\}$-Einstein equation, is
\begin{equation}
\label{energy}
E_{{\cal L}}=6\,f''(R)\,a^2\,\dot a\,\dot R+ 6\,f'(R)\,a\,\dot
a^2- a^3\,\left[f(R)-f'(R)\,R\right] +6k\,f'(R)\,a-D=0\, .
\end{equation}
The equations of motion for $a$ and $R$ are respectively
\begin{equation}\label{eqa}
f''(R)\left[R+6\,H^2+6\,\frac{\ddot a}a+6\,\frac{k}{a^2}\right]=0
\end{equation}
\begin{equation}\label{eqR}
6\,f'''(R)\,\dot R^2+ 6\,f''(R)\,\ddot R+
6\,f'(R)\,H^2+12\,f'(R)\,\frac{\ddot a}a=
3\,\left[f(R)-f'(R)\,R\right]-12\,f''(R)\,H\,\dot R -6\,f'(R)\,\frac{k}{a^2}\, ,
\end{equation}
where $H\equiv \dot a/a$ is the Hubble parameter. Eq.~(\ref{eqa})
ensures the consistency, since $R$ coincides with the definition
of the Ricci scalar in the FRW metric.

The choice\footnote{The reason for the absolute value stays only
in the fact that, with our conventions, $R$ terms out to be
negative. It is obviously possible to rewrite everything with
$f(R)=R^ {3/2}$ and $R>0$.} $f(R)=-|R|^{3/2}$ in
Eqs.~(\ref{Ag}-\ref{eqR}) produces a theory able to describe dust
matter and dark energy combined phases in a FRW spacetime, without the need of any extra field
introduced {\it ad hoc} (see \cite{Capozziello:2008im,Astronf(R)} for details). In this
particular case, the point-like FRW Lagrangian (\ref{pL}) is
\begin{equation}
{\cal L}=\frac{a^3}{2}|R|^{3/2}-\frac{9}{2}a^2|R|^{-1/2}\dot R\,\dot a+9\,|R|^{1/2}a\,\dot a^2-9\,k|R|^{1/2}a+D\, ,
\end{equation}
and the energy function
\begin{equation}\label{pLp}
E_{{\cal L}}=-\frac{9}{2}a^2|R|^{-1/2}\dot R\,\dot a+9\,|R|^{1/2}a\,\dot a^2-\frac{a^3}{2}|R|^{3/2}+9\,k|R|^{1/2}a-D=0\, .
\end{equation}
Referring to \cite{Capozziello:2008im} , it is possible to show that such a model has 
a Noether symmetry that allows 
 to find out an exact solution for Eqs.(\ref{energy}), (\ref{eqa}) and
(\ref{eqR}) for this particular $f(R)$, that is
\begin{equation}\label{ag}
a(t)=\sqrt{a_4 t^4+a_3 t^3+a_2 t^2+a_1 t}.
\end{equation}
with
\begin{equation}\label{p}
a_{4}=\frac{\Sigma_{1}^{2}}{144}\quad;\quad a_{3}=\frac{\Sigma_{1}\Sigma_{0}%
}{36}\quad;\quad a_{2}=\frac{\Sigma_{0}^{2}}{24}-k\quad;\quad a_{1}%
=\frac{\Sigma_{0}^{3}}{36\Sigma_{1}}-2k\frac{\Sigma_{0}}{\Sigma_{1}}+\frac
{4D}{9\Sigma_{1}}.
\end{equation}
where $k$ is the spatial curvature, $\Sigma_1$ the Noether charge
and $\Sigma_0$ the integration constant.

In order to fix the coefficients $a_{i}^{\prime }s$, we have to consider 
time units in which the current time is $t_{0}=1$.  However, one
can construct the dimensionless quantity $H_0 t_0 \sim 0.93$ which
has to remain constat. Therefore the Hubble parameter results of order
one, (we choose $H_0=1$ for simplicity). The current deceleration
parameter can also be fixed taking $q_0=-0.4$, which could
describe a reasonable current acceleration. Finally, a unit value
for the present scale factor value  is considered. This assumption can be
always done if no restriction on the value of $k$ is imposed. In
order to fix the remaining free parameters, we consider
$a_4=0.106$, which leads $\Omega_{m0}=0.0418032$ (with
$\Omega_m=\rho/[6H^2f'(R)]$), very close to the expected content
of baryonic matter. With these assuptions, the scale factor is
\begin{equation}\label{aJ}
a(t)=\sqrt{\frac{t}{5}[2+0.53(t-1)^3+t+2t^2]}
\end{equation}
and the Ricci scalar
\begin{equation}\label{RJ}
R(t)=\frac{9(41+212t)^2}{212t(147+259t+41t^2+53t^3)}\, .
\end{equation}
This model describes a spatially open universe, $k\simeq-0.5$. We have
to note that the measurable quantity is not this parameter but
$\Omega_{k0}\simeq 0.02$ which is very small. Moreover, since the
requirement $\Omega_k\simeq 0$ is derived by the spectrum of the
CMBR  data, and these data strongly depend on the
standard $\Lambda$CDM model, we cannot conclude that this
feature is needed in our $f(R)$-model.

In fact, this solution, in principle, seems to reproduce
satisfactorily  observational data, out from the trivial
fulfilment of the {\it a priori} fixed. In particular, the scale
factor (\ref{aJ}) is able to emulate a dust dominated epoch
necessary for the structure formation, with only a difference with
respect the standard $a_F\sim t^{2/3}$ of the 3 $\%$ in the
range $2\leq z\leq 4$, and the distance modulus derived by this
model is also able to reproduce the SNeIa data,
\cite{Capozziello:2008im}\footnote{This choice of the parameters
is interesting because it produces results which turn out to be
reasonably good at least from the point of view of observational tests. However,  the
following comparison with the Einstein frame is not dependent on
this choice}.

\section{Conformal transformation}\label{dos}

Let us consider now  the gravitational part of our action, i. e.
\begin{equation}
{\cal A}_G=-\int d^4 x\,\sqrt{-g}\,|R|^{3/2}\, ,
\end{equation}
which, by defining a auxiliary scalar field $\varphi$ in the
following way,
\begin{equation}\label{campoR}
\varphi(R)=\sqrt{\frac{3}{2}}\ln\left(3|R|^{1/2}\right)\, ,
\end{equation}
can be written as
\begin{equation}\label{auxiliar}
{\cal A}_G=\int d^4
x\,\sqrt{-g}\,\left[-\frac{|R|}{2}e^{\sqrt{2/3}\varphi}+\frac{1}{54}e^{3\sqrt{2/3}\varphi}\right]\,
.
\end{equation}
The new field $\varphi$ does not introduce any physical new
feature, since it is only a way to recast the further gravitational degrees of freedom related to $f(R)$-gravity. In
fact, it can be seen that this is the case, since the
$\varphi-$field equation obtained from Eq.~(\ref{auxiliar})
produces only Eq.~(\ref{campoR}). If we perform a conformal
transformation by the conformal parameter
\begin{equation}\label{bvarphi}
b(t)=\exp\left(\frac{\varphi}{2}\sqrt{\frac{2}{3}}\right)\, ,
\end{equation}
which is a function of the time $t$ since
$\varphi(R(t))=\varphi(t)$, the resulting action is the
Hilbert-Einstein action with a scalar field $\varphi(t)$
\begin{equation}
{\cal A}_G=\int d^4
x\,\sqrt{-\bar{g}}\,\left[-\frac{|\bar{R}|}{2}-
\frac{1}{2}\bar{g}^{\mu\nu}\partial_\mu\varphi\partial_\nu\varphi+V(\varphi)\right]\,,
\end{equation}
where $||\bar{g}_{\mu\nu}||=b(t)^2{\rm
diag}(-1,a(t)^2,a(t)^2,a(t)^2)$, $\bar{R}$ is the Ricci scalar of
the metric $\bar{g}_{\mu\nu}$ and $V(\varphi)=\exp [\sqrt{2/3}\varphi]/54$. If we define a new time variable
$\tau$, in such a way that $d\tau=b(t)d t$, we recover a FRW
metric $\tilde{g}_{\mu\nu}$, but now with a scale factor
$a_E(\tau)=b(\tau)a(\tau)$
\begin{equation}
{\cal A}_G=\int d^4
\tilde{x}\,\sqrt{-\tilde{g}}\,\left[-\frac{|\tilde{R}|}{2}-
\frac{1}{2}\tilde{g}^{\mu\nu}\tilde{\partial}_\mu\tilde{\varphi}\tilde{\partial}_\nu\tilde{\varphi}+
\tilde{V}(\tilde{\varphi})\right]\,
\end{equation}
$\tilde{R}$ is the Ricci scalar of the metric
$\tilde{g}_{\mu\nu}$, $\tilde{R}(\tau)=\bar{R}(t(\tau))$,
$\tilde{\varphi}(\tau)=\varphi(t(\tau))$ and
$\tilde{V}(\tilde{\varphi})=V(\varphi)$. Taking also into account
the mentioned transformations in the matter component, we obtain
the total action in the Einstein frame and the point-like FRW
Lagrangian
\begin{equation}\label{AE}
{\cal L}=3\,a_E\left(\partial_\tau a_E\right)^2-3\,k\,a_E-\frac{a_E^3}{2}\left(\partial_\tau\tilde{\varphi}\right)^2+
a_E^3\tilde{V}(\tilde{\varphi})+e^{-\tilde{\varphi}/\sqrt{6}}\tilde{\rho}_{m}\, ,
\end{equation}
where $\tilde{\rho}_{m}=D/a_E^3$. Such a  Lagrangian shows a coupling
between the matter term and the scalar field, which will produce
the non-conservation of both fluids individually.

The Einstein equations yield
\begin{equation}\label{eqeinsteinE}
\tilde{G}_{\mu\nu}=\tilde{T}_{\mu\nu}^{\tilde{\varphi}}+\tilde{T}_{\mu\nu}^{m}+\tilde{T}_{\mu\nu}^{int}\,
,
\end{equation}
where
\begin{equation}\label{Tvarphi}
\tilde{T}_{\mu\nu}^{\tilde{\varphi}}=\tilde{\partial_\mu}\tilde{\varphi}\tilde{\partial_\nu}\tilde{\varphi}-
\frac{1}{2}\tilde{\partial_\alpha}\tilde{\varphi}\tilde{\partial^\alpha}\tilde{\varphi}\tilde{g}_{\mu\nu}+
\tilde{V}(\tilde{\varphi})\tilde{g}_{\mu\nu}
\end{equation}
\begin{equation}
\tilde{T}_{\mu\nu}^m={\rm diag}(\tilde{\rho}_{m},0,0,0)\,  ,
\end{equation}
and
\begin{equation}
\tilde{T}_{\mu\nu}^{int}=\left(e^{-\tilde{\varphi}/\sqrt{6}}-1\right){\rm
diag}\left(\tilde{\rho}_m,0,0,0\right)\,.
\end{equation}
It should be noted that, whereas $\tilde{T}_{\mu\nu}^m$ is conserved
$\tilde{T}_{\mu\nu}^{\tilde{\varphi}}$ and
$\tilde{T}_{\mu\nu}^{int}$ do not fulfil any conservation law
separately, but
$\left(\tilde{T}_{\mu\nu}^{\tilde{\varphi}}+\tilde{T}_{\mu\nu}^{int}\right)^{;\mu}=0$.
This result has to be taken into account in order to compare results in Jordan and Einstein frames.

\section{Jordan frame versus Einstein frame}\label{tres}
In the previous section, we have shown how to
perform a conformal transformation of  $f(R)$-gravity to obtain 
GR  with a dynamical scalar field, being therefore both
frames mathematically equivalent. However, this mathematical
equivalence does not necessary ensure the physically equivalence
of both frames. In fact,  whereas, in
the Jordan frame, the matter term is not-coupled to any field or to
gravity, in the Einstein frame there is a coupling between the
matter and the scalar field, appearing as an interaction term in
the Einstein equations (\ref{eqeinsteinE}) . This fact is crucial in comparing the physics in the two systems.

In order to show that the two frames could be
physically equivalent, we have to compare the physical quantities of
the mentioned two frames. This is a delicate issue since the selection of such quantities should be unambiguous. 

Through the definition of the conformal
factor, Eq.~(\ref{bvarphi}), and Eqs.~(\ref{RJ}) and
(\ref{campoR}), one finds the explicit form of this parameter in
terms of $t$
\begin{equation}\label{b}
b(t)=\frac{3\sqrt{41+212t}}{\sqrt{106}\left(147t+259t^2+41t^3+53t^4\right)^{1/4}}\, ,
\end{equation}
with $t$ the cosmic time in the Jordan frame, which is related to the cosmic time in the Einstein frame
\begin{equation}\label{taut}
\tau=\int b(t)\,d\,t\, .
\end{equation}
Since $a_E(t)=b(t)a(t)$, Eq.~(\ref{b})  allows to obtain the
scale factor in the Einstein frame in terms of $t$ and, therefore,
in terms of $\tau$ trough Eq.~(\ref{taut}). In such a way, taking
into account Eqs.~(\ref{bvarphi}) and (\ref{taut}), one can
known, in principle,  the explicit form of $\tilde{\varphi}(\tau)$. Unfortunately,
it is not possible to obtain an analytic solution for
$\tau(t)$, but we can perform a complete analytic study in terms
of $t$, noting that, in the Einstein frame, it is only an arbitrary
parameter and not the cosmic time. We thus maintain the dot for
derivation with respect to $t$ and write explicitly the
derivatives w.r.t. the cosmic time $\tau$. This procedure will not
affect the final results, because they will be set in terms of the
redshift, which is an observable quantity.

\begin{figure}[t]
\centering \resizebox{10cm}{!}{\includegraphics{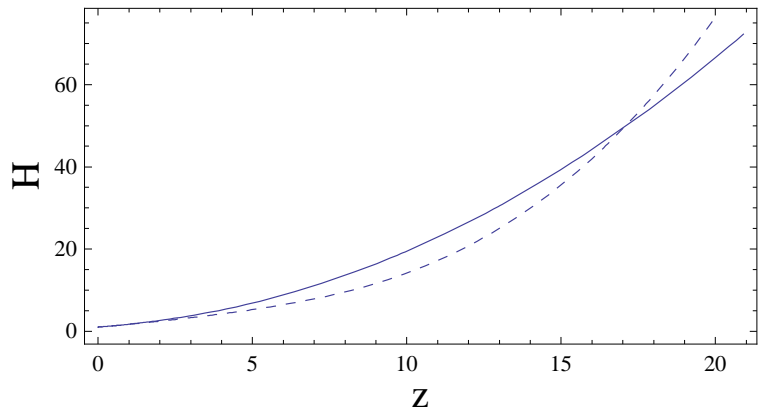}}
\caption{Comparison of the Hubble parameter, $H(z)$ in the Jordan
frame and in the Einstein frame (dashed line), where the Hubble
parameter in the Einstein frame has been normalized with its
current value.} \label{figura1}
\end{figure}

\begin{figure}[t]
\centering \resizebox{10cm}{!}{\includegraphics{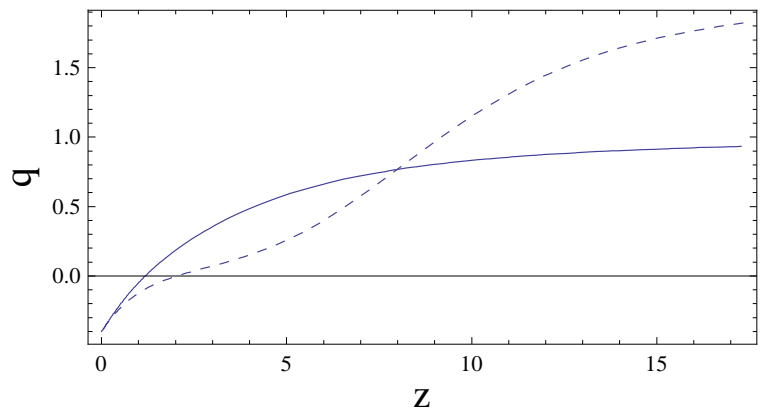}}
\caption{Comparison of the deceleration parameter, $q(z)$ in the
Jordan frame and in the Einstein frame (dashed line).}
\label{figura2}
\end{figure}

Taking into account that $a_E(t)=b(t)\,a(t)$, we get the Hubble
parameter in the Einstein frame
\begin{equation}\label{HE}
H_E(t)=\frac{\partial_\tau a_E}{a_E}=\frac{1}{b(t)}\frac{\dot{a}_E}{a_E}\, ,
\end{equation}
and a deceleration factor
\begin{equation}\label{qE}
q_E(t)=-\frac{\left(\partial_\tau^2a_E\right)a_E}{\left(\partial_\tau
a_E\right)^2}=
-\frac{\ddot{a}_E\,a_E}{\dot{a}_E^2}+\frac{\dot{b}\,a_E}{b\,\dot{a}_E}\, .
\end{equation}
Since the redshift can also be defined in terms of the
parameter $t$,
\begin{equation}\label{zE}
z_E(t)=-1+\frac{a_{E,0}}{a_E(t)}\, ,
\end{equation}
where $a_{E,0}$ is the current scale factor, we can eliminate the
(unphysical) parameter $t$, by considering couples of parametric
equations. In order to perform this study, we must fit
$t_0=t(\tau_0)$, and we do that demanding that the dimensionless
parameter $q_{E,0}=-0.4$ as it was required in the Jordan frame,
setting the value $t_0\simeq 1.24$. 
Figs.~\ref{figura1} and \ref{figura2} show that the Hubble
parameter $H(z)$ and the deceleration parameter $q(z)$,
respectively, are different in the Jordan and Einstein frames.
This means that the frames are not physically equivalent (in
fact, it would be enough that one of these physical functions were
different in the two frames).

One can also compare the dimensionless quantity $\Omega_{m,0}$ in
both frames. In the Jordan frame, one can easily see, from the
$00$-component of Eq.~(\ref{eqeinsteinJ}), that it must be defined
as $\Omega_{m,0}=\rho_{m,0}/(6f'(R)\,H_0^2)$ and takes a value
compatible with the baryonic component of the Universe, i. e.,
around $0.04$. This parameter is defined in the Einstein frame as
$\tilde{\Omega}_{m,0}=\tilde{\rho}_{m,0}/(3H_{E,0}^2)$, and takes
a value which is more than twice the value in the Einstein frame,
that is $\tilde{\Omega}_{m,0}\simeq 0.09$. On the other hand, in
the Einstein frame there is an interaction term which produces
$\tilde{\Omega}_{int,0}=(1/b-1)\tilde{\rho}_{m,0}/(3H_{E,0}^2)=-0.0567$,
therefore its absolute value is more than one half the value of
the matter component, so it should produce some observable effect.

\begin{figure}[t]
\centering \resizebox{10cm}{!}{\includegraphics{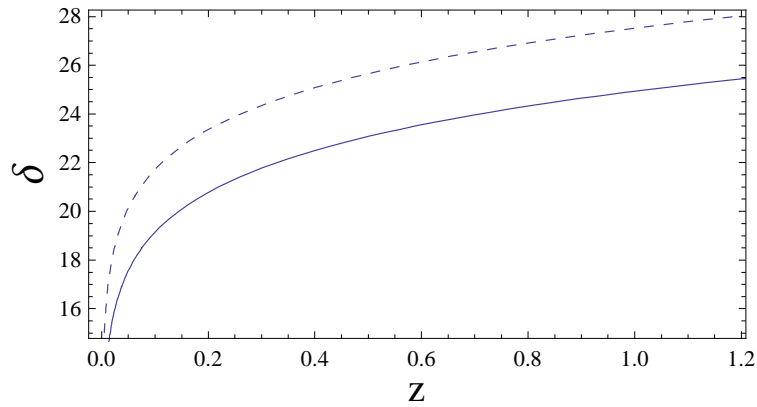}}
\caption{Comparison of the distance modulus in the Jordan frame
and in the Einstein frame (dashed line).} \label{figura3}
\end{figure}

In order to show even more clearly than the Jordan and Einstein
frames are not equivalent, we illustrate this fact in the
following way. Let us consider two different researchers studying
the model presented in sec.~\ref{uno} following two different
routes. One of them refers all its calculations to the original
Jordan frame and conclude that this model can describe the
distance modulus data, as it is shown in
\cite{Capozziello:2008im}. The other one considers that the Jordan
frame and the Einstein frame are physically equivalent and
calculate also the distance modulus, but in the Einstein frame. As
it is shown in Fig.~\ref{figura3} , they obtain
different functions. Since the function calculated in the Jordan
frame fits the mentioned data, while the function
obtained in the Einstein frame does not, the second research would
conclude that the model does not describe our Universe, whereas the
first one would continue with his study.

\section{Conclusions }\label{cuatro}
In this letter,  we have shown that the Jordan and Einstein frames could not be
physically equivalent according to the choice of observable quantities. We have consider a particular
$f(R)$-model and the resulting model in the Einstein frame,
obtained by a conformal transformation. The discrepancy between
these models has clearly been shown in the coupling term between
the matter component and the scalar field which appears in the
conformally transformed model in the Einstein frame, and in
Figs.~\ref{figura1}, \ref{figura2} and \ref{figura3}, which
prevent that the two models could describe the "same" Universe.
These differences cannot be considered as a mistake coming from
any numerical approximation, since all the study is performed in
an analytic way.

On the other hand, 
conformal transformations between the Jordan and Einstein frames result extremely useful
if used in a consistent way. Thus, since the Jordan and
Einstein frame are  mathematically equivalent, one can
perform the calculations in the more convenient frame whenever one
conformally transforms the obtained functions to the ``true frame''.

The identification of  the "true" physical  frame  is a 
controversial question. But if one consider that the Jordan
(Einstein) frame is the true frame, one must refer all results to
this frame in order to compare them with the observational data.
One can also take an equitable position and consider that the "true
frame" is that which is in agreement  with the observational data to a
larger extent. This point remain still open, although the
 model presented here  and in  \cite{Capozziello:2008im} is  able to fulfil some
observational tests (without the introduction of any dark stuff) and
to reproduce a dust matter decelerated phase, before the current
accelerated one, may  help us to find an answer in a future.
Obviously a deeper study of the mentioned model is still
necessary and the physical non-equivalence between frames should be tested also for other models and observables.

\section*{Acknowledgements}
PMM gratefully acknowledges the financial support provided by the
I3P framework of CSIC and the European Social Fund.

\end{document}